\documentclass[fleqn,twoside]{article}
\usepackage{espcrc2}
\usepackage{longtable}
\usepackage{graphicx}
\title{Late Stages of Stellar Evolution}
\author{Kevin Marvel
\address[AAS]{American Astronomical Society\\
2000 Florida Avenue, NW \\
Suite 400 \\
Washington, DC 20009 \\
USA}}
\begin{document}
\begin{abstract}
The Square Kilometer Array will have the sensitivity, spatial
resolution, and frequency resolution to provide new scientific
knowledge of evolved stars.  Four basic areas of scientific
exploration are enhanced by the construction of the SKA: 1) detection
and imaging of photospheric radio continuum emission and position
correlation with maser distributions, 2) imaging of thermal dust
emission around evolved stars and the detailed structures of their
circumstellar winds (again, including comparison with maser
distributions), 3) study of cm-wavelength molecular line transitions
and the circumstellar chemistry around both O-rich and C-rich evolved
stars and 4) the possible observation of polarized emission due to the
influence of the magnetic fields of AGB stars.  Since this short
chapter is not meant to be a review article, a comprehensive reference
list has not been generated.  I have selected just one or perhaps two
references for citations where appropriate.
\vspace{1pc}
\end{abstract}
%
\maketitle
\section{Introduction}

Although much can be learned by studying stellar nurseries and the
fascinating process of stellar birth, we have much yet to learn in the
field of stellar geriatrics.  Stars that do not proceed to explosive
ends, the low- and intermediate-mass stars, undergo a period of
mass loss, often extreme, in which 50\% or more of the star's initial
mass is transferred back to the ISM.  The rates of this mass loss vary
widely from $10^{-6} \/ \mbox{M}_\odot$ per year to as much as
$10^{-4} \/ \mbox{M}_\odot$ per year.

Such prodigious mass loss and the large number of low- and intermediate
mass stars results in the fact that most of the interstellar
medium---perhaps as much as 80 to 90 percent---has been cycled through
a star and ejected via this process.  Flash-in-the-pan supernovae do
have a significant impact, especially enriching the heavier metals in
the ISM, but the bulk of the material is provided by the aging process
of common stars similar to the Sun.  Understanding how the mass loss
process proceeds and its implications on the chemical modification of
the ISM in our own galaxy has obvious implications for the study of
more distant galaxies as well as being of interest itself.

The mass loss process proceeds from the formation of dust in the upper
atmospheres of evolved stars, a few stellar radii from the optical
photosphere.  This process has long been thought to be driven by the
pulsations inherent in these kinds of stars, but it now appears likely
that it is driven by dramatic temperature fluctuations caused by the
formation of TiO in the stellar atmosphere, which changes the physical
conditions in the dust formation region \cite{Reid2003}.

Although the details of dust formation remain an unknown factor, we
know roughly that when the temperature and density conditions are
appropriate, nucleation can occur, leading to the formation of dust.
This dust, exposed to the radiation field of the evolved star, absorbs
outward momentum and begins to accelerate. Gas not incorporated into
dust grains is carried along with the dust through momentum coupling.
As conditions allow, molecules can form from the gas that is carried
along with the outward-moving dust.  Using existing centimeter and
millimeter wave interferometers, studies of the molecules formed in
these winds have been completed showing more or less spherically
symmetric mass loss \cite{Rieu1990} with some interesting results such
as rotation \cite{Bieging1996} and perfect spherical symmetry of an
apparently single ejection event \cite{Olofsson1998}.

Certain molecules are capable of maser emission (e.g. SiO, H$_2$O and
OH).  When such masers are found in the winds of evolved O-rich stars,
they are powerful probes of the mass loss kinematics.  Some C-rich
stars do exhibit HCN masers at high frequencies, but the O-dominated
species are absent.  However, they can provide only rough information
about the physical conditions of the wind itself, provided by the
physical conditions required for maser emission.  Using VLBI
techniques, which provide resolutions as fine as 100 $\mu$arcseconds,
the motions of masing gas can be tracked with high accuracy and the
kinematics of the wind modelled.  In practice, this has proven to be a
challenging undertaking.  The non-linear emission process and
apparently complex distributions of the masers make modelling
difficult.  Only rough models have yet been made placing the masers in
ellipsoidal distributions undergoing a variety of kinematic motions.
The maser observations do indicate more-or-less spherically or
elliptically symmetric mass loss with acceleration occurring to the
outermost regions of the wind where acceleration ceases due to decoupling
of the gas from the dust.  A mild controversy about the relative
angular scales of the OH and H$_{2}$O maser distributions (e.g. the OH
masers, although predicted to be at large radii, appear at about the
same angular scale as the H$_{2}$O masers) is likely due to beaming
effects.  Water masers are preferentially tangentially beamed as they
reside in an accelerating portion of the wind while the OH masers are
radially beamed as they reside in a constant velocity region of the
wind \cite{ReidMASERMTG}.  As difficult as the physics and geometries
are, our current understanding is limited due to lack of adequate
modelling in my opinion.  Other results indicate non-negligible
rotation \cite{Boboltz2000} of the envelopes and the influence of
magnetic fields on the shape of the shell \cite{Murakawa2003}.

We have yet to understand the dust formation process in these objects.
Although infrared interferometric observations \cite{Monnier2004} hint
at a very clumpy and dynamic process, we have few tools available to
probe this process in detail.  We have only a very rough picture of
the structure of the extended photosphere and wind of evolved stars.
The role of magnetic fields in AGB stars has not been explored in any
detail, though they must impact the dust formation process, the wind
itself and obviously provide information on the star itself.

Although detailed studies have been made using millimeter
interferometers of the chemical structure of the nearest and largest
evolved stars, much of the chemical structure in these objects remains
a mystery.  ALMA will help here but will miss the low-frequency line
transitions.  A detailed understanding of the structure of these
objects awaits the Square Kilometer Array.

The SKA, with its high resolution, sensitivity to a range of emission
mechanisms and low-frequency observing capability will allow studies
of evolved stars that have not been possible before and provide
complementary observations to those provided by ALMA and other
instruments.  I discuss the anticipated observations SKA can provide
in the sections below.

\section{Imaging the Surfaces of AGB Stars}

\subsection{Fluctuations First}

It has been shown \cite{Petit1933} that certain AGB stars, the Mira
variables, undergo temperature changes of 30\% and luminosity changes
of a factor of two during their visual fluctuation period ($L=\sigma
T^{4}_{e} \pi R^{2}$).  As shown in \cite{Reid2003}, these changes
should result in stellar radius fluctuations of about 40\%.  Such
dramatic fluctuations would lead to both dramatic shock waves that
propagate from the star into its extended photosphere and also
measurable changes in light curves at radio, infrared and optical
wavelengths.  The visual light curve fluctuates dramatically (extreme
cases show fluctuations of 8 magnitudes) while the infrared light
curves rarely fluctuate by more than a magnitude and the radio light
curves fluctuate only by a few percent at most.

\begin{figure*}
\includegraphics*[width=16cm]{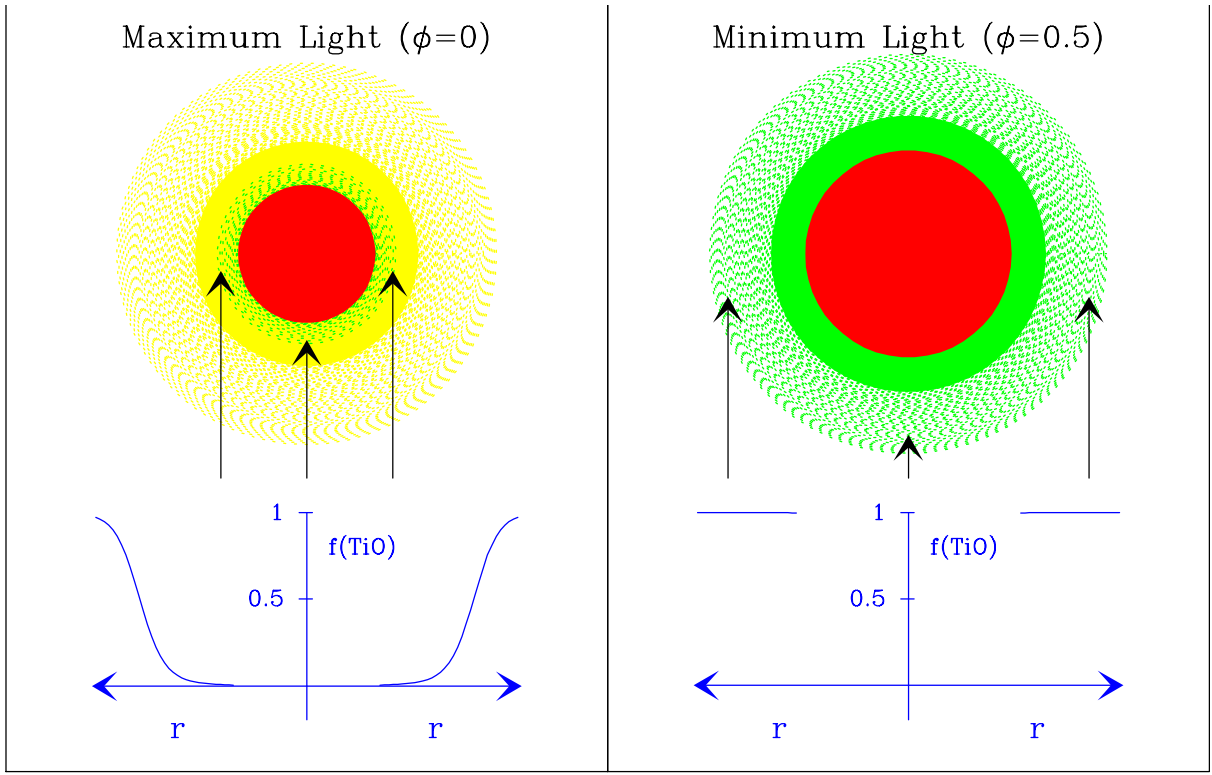} \caption{A schematic
depiction of the change in visual appearance of a Mira variable star
at maximum ({\it left-hand panel}) and minimum ({\it right-hand
panel}) light.  The star, shown in {\it red,} is smaller and hotter at
maximum light than at minimum light.  At maximum light, the extended
atmosphere of the star (shown as {\it yellow}) is partially
transparent at visual wavelengths, and one sees almost down to the
stellar surface (indicated with {\it arrows}).  Near minimum light,
the temperature of the star has declined and metallic oxides, such as
TiO (shown as {\it green}), form throughout the extended atmosphere.
The fraction of Ti in TiO, f(TiO), as a function of radius is plotted
in {\it blue}.  Near minimum light, TiO forms with sufficient density
at a radius of $\approx1.8R_*$ to become opaque to visible light.  At
this radius, the temperature can be very low, and almost all radiation
is in the infrared.  Since little visible light emerges, the star can
almost disappear to the human eye.  Figure and caption text taken from
\cite{Reid2003}. \label{REIDMIRA}} 
\end{figure*}

Observed light curves do not match those predicted by the radial
fluctuations implied by the temperature and luminosity fluctuations
\cite{Reid2003}.  Figure \ref{REIDMIRA} shows the model from
\cite{Reid2003} that (to first order) reproduces the radio, infrared
and visual light curves.  The model predicts that TiO is formed in the
upper atmosphere as the star approaches minimum light and this
additional opacity source can greatly decrease the observed light at
visual wavelengths while having less impact at infrared wavelengths
and almost no impact at radio wavelengths.  This new discovery shows
that much remains to be learned about evolved stars.  After all, Mira
variables are one of the oldest astronomical phenomena studied and
only now has an adequate first-order model been developed to explain
their fluctuations.  


The SKA will allow further testing of these models at far greater
sensitivity.  Observations of the flux from the photospheres of AGB
stars are exceedingly difficult.  Typical fluxes are on the order of
200 microJy and require special calibration techniques with current
interferometers.  The SKA, with a sensitivity of about 0.1 microJy at 20
GHz will provide the most accurate AGB star light curves across
all wavelengths.  Such measurements will allow improved modelling of
the opacity source fluctuations in the star.

The discovery of particularly large extrasolar planets orbiting close
to their host star opens up the possibility for the observation of
eclipses using the SKA, as has been observed in the star HD 209458.
However, the eclipse type that is potentially observable would be an
active radio-emitting planet similar to Jupiter being eclipsed by its
host star rather than the more typical eclipse.  As pointed out in
\cite{SKA1}, Jupiter-like planets will produce detectable radio
emission out to distances of 10 pc.  The passage of a planet of this
type behind its host AGB star would be detectable, since the emission from
the planet would be of order 10 microJy, compared to the photospheric
flux of 200 microJy.

\subsection{Imaging Second}

The diameter of the radio photosphere of a typical AGB star is of
order 5 AU.  At 1kpc, such a source would have a maximum angular
diameter of 6 milliarcseconds.  At 22 GHz and with 1000 km baselines,
the SKA will have a resolution of roughly 3 milliarcseconds.  This
resolution corresponds to linear resolution of 3 AU at 1 kpc.  For AGB
star diameters of 3-5 AU, they can be moderately resolved with SKA.
Thus, for only the nearest AGB stars will any degree of imaging be
possible.  The number of AGB stars closer than 1kpc is limited.
Without a substantial increase in the highest frequency
observed by the SKA or the maximum baselines, imaging of only the
nearest AGB stars will be possible.

That said, some very interesting imaging projects can be undertaken
for large AGB stars not further than 1 kpc from the Earth.  For
example, the well-known and nearby (150 pc) carbon star IRC+10216
has a photospheric size of 35 milliarcseconds and an extended envelope
diameter of nearly 1'.  With a resolution element of 3 mas, the
surface of the star would be imaged well and the overall envelope,
especially in spectral lines (see below) would be highly resolved.
The imaging design goal of 0.1 arcsecond resolution at 1.4 GHz over a
1 degree field (and scaled with frequency) is sufficient to provide
high spatial dynamic range imaging at high sensitivity for objects of
this type.  The science the SKA will allow is the direct imaging of
the dust formation process and connection with stellar pulsation for
the nearest and largest AGB stars.

\section{Observations of Masers and their Host Stars}

Maser emission from gas in the outflowing winds of AGB stars is a
common phenomenon in O-rich AGB stars.  Masers are regions of gas in
the stellar wind that have sufficient velocity coherence to amplify
background photons via amplified emission of radiation.  Such
amplification is possible due to a population inversion of the
molecular species in question and a fortuitous alignment of
molecular rotational, vibrational or ro-vibrational energy levels.
Several species are found.  SiO masers are located close to the star
(within a few stellar radii and below the dust formation zone).
H$_{2}$O masers are located at intermediate distances from a few tens
of stellar radii to a few hundred.  The remote OH masers are located
up to several thousand stellar radii from the host star.

\begin{figure}
\includegraphics[width=7.5cm]{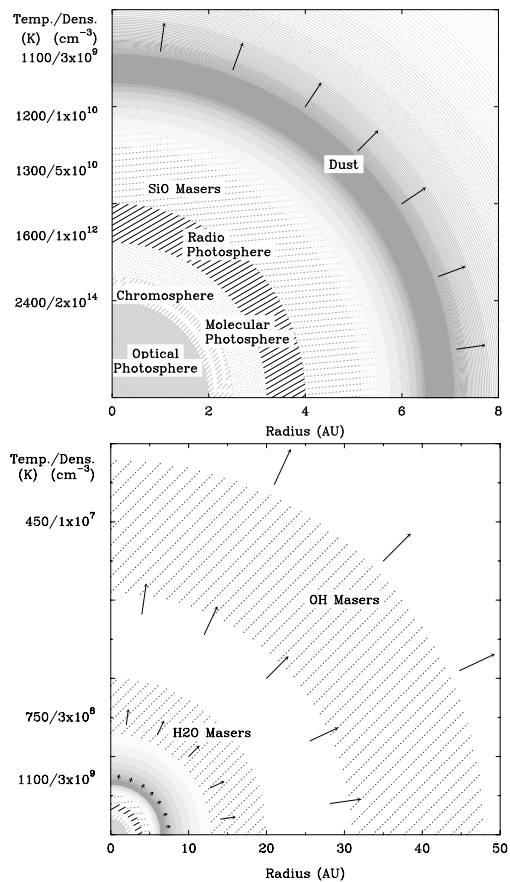}
\caption{
A schematic showing our current understanding of the circumstellar
region around an AGB star \cite{ReidMASERMTG}.
\label{SHELL}
}
\end{figure}

In addition to knowing the location of the various maser species, we
have a good understanding of the overall shell structure around these
stars \cite{ReidMASERMTG}.  Figure \ref{SHELL} shows graphically our
current understanding of the circumstellar region around an AGB star.
The star itself is from between 1 and 5 AU in size.  Above this
surface is a chromospheric region followed by a molecular photosphere
ending between 1 and 2 AU above the optical photosphere.  The radio
photosphere (about 0.5 to 1AU in thickness) is located near the SiO maser
formation region.  Beyond this zone, wind acceleration begins as dust
forms in a region from roughly 5 AU to 10 AU (depending on the
properties of the star and pulsation phase).  The H$_{2}$O and OH
masers begin to appear at radii of 15 AU or more and the OH masers are
found further out from the water maser shell.  The exact sizes of the
various regions, their exact locations and how they interact remain
rough measurements.

Masers are imaged using VLBI techniques and typically have resolved
sizes of a milliarcsecond or so, but observations with MERLIN show
that a weak diffuse emission can also be present \cite{Richards1999}.
Depending on the upper frequency cutoff for the SKA, the OH (1.6 GHz),
methanol (6.7 GHz) and water masers (22 GHz) could be observable.
However, it is not the detection of maser emission with the SKA that
is of greatest interest (though the sensitivity of the instrument
would allow detection of extragalactic masers to a much greater
distance than currently available).  It is the sensitivity to both the
stellar photospheric emission and the dust continuum in the wind
combined with the VLBI observations that will be of prime interest.

\begin{figure*}
\includegraphics*[width=16cm,angle=90]{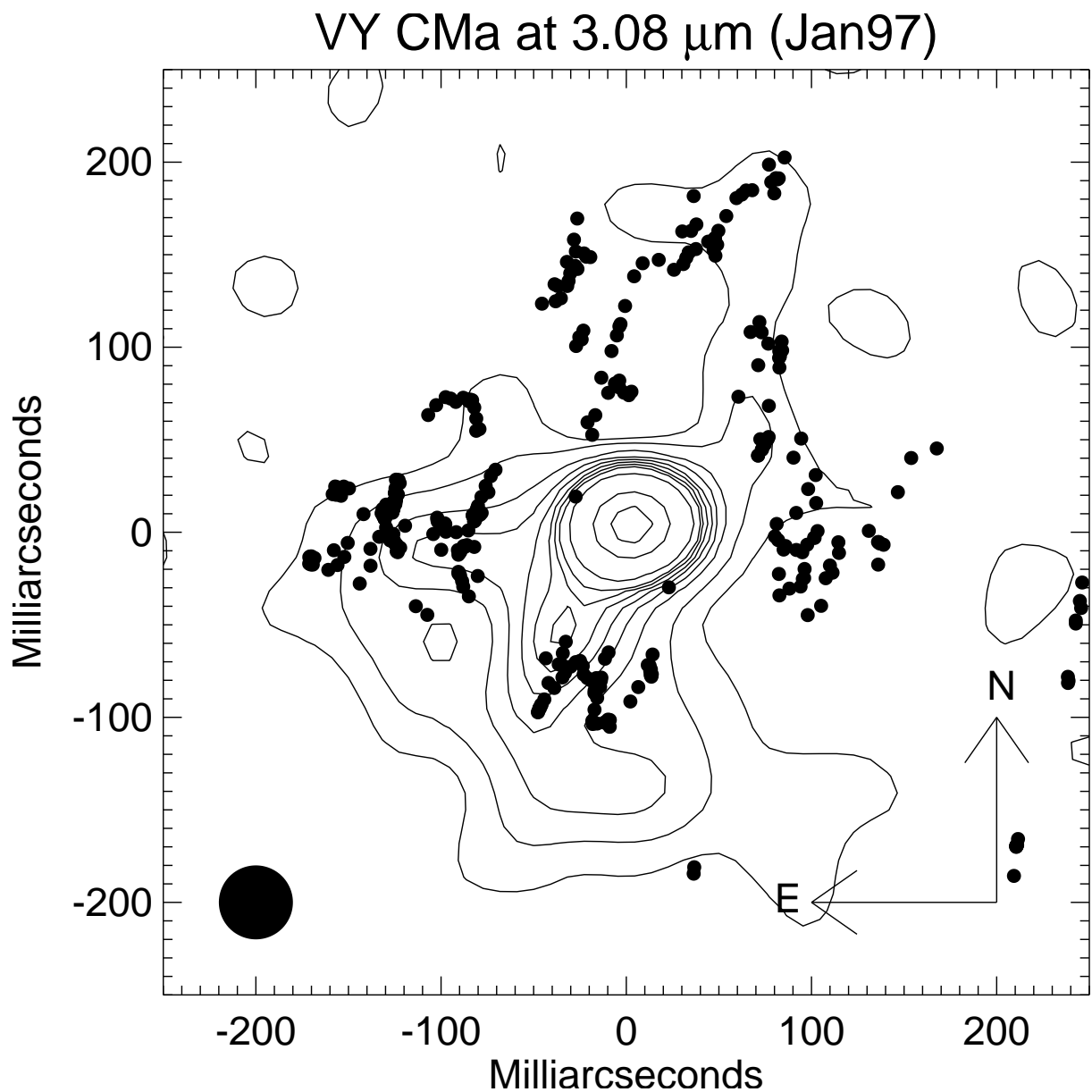} \caption{This figure shows
the locations of water masers detected with MERLIN (note beamsize of 40 mas in
lower left hand corner of figure) overlaid on infrared emission at 3.08 $\mu$m.
The infrared emission was imaged using an interferometric masking technique
on the Keck 10m telescope.  Although registration of the images is a challenge
technically, the image shows the power of such combination observations.  Multi-epoch
observations of both the water masers and the infrared emission shows changes over time
and there is some hope of making time-lapse movies of these sources in the future.
Figure courtesy of J. Monnier.
\cite{MonnierFigure}, \cite{RichardsFigure}. \label{MonnierFigure}} 
\end{figure*}

VLBI imaging techniques are sensitive only to very high brightness
temperatures and the smallest angular sizes ( ~ 1-5 milliarcseconds)
and therefore only the maser spots themselves and not the environment
in which they are located can be imaged.  With the angular resolution
of the SKA at 1.4 GHz (0.1"), the thermal emission across a typical OH
maser distribution 1"-2" in diameter could be mapped with sufficient
resolution and sensitivity to allow alignment of the VLBI maser
observations with the overall dust distribution and star itself.
Combined with infrared interferometric observations, which are now
beginning to show the details of the dust distribution at high angular
resolutions (see Figure \ref{MonnierFigure}) \cite{Monnier2004} ($\approx 
10$ milliarseconds, but over limited
fields of view), the SKA will play a critical role in providing
information on the largest scales.

Outstanding problems to be addressed include the details of dust
formation, such as whether the process proceeds uniformly as a
function of pulsation cycle or at particular times, the degree of
clumpiness of the dust formation and the exact physical conditions
that lead to dust formation.  The transition of AGB stars from roughly
spherically symmetric mass-losing objects to the asymmetric planetary
nebulae has yet to be understood completely and the combination of the
kinematic information provided by VLBI maser observations and the dust
distribution will hopefully shed new light on this area.

\section{Molecular Gas}

In the frequency range of the SKA are 634 molecular line transitions,
many of which have not been well studied, only detected, and some of
which have still not been identified \cite{LOVAS}.  For convenience,
these transitions are provided in Table \ref{LOVASTABLE} .

Many of these species are expected to be present in the winds of
evolved stars.  As pointed out by Zijlstra (2003), both the dust and
gas created in the stellar winds of AGB stars survive in the ISM.  The
dust, as indicated by reddening and particles found in meteorites or
as micrometeoroid particulates in our own upper atmosphere
\cite{Messenger2003}, survives in interstellar space.  The presence of
the Diffuse Interstellar Bands are the main piece of evidence for the
existence of rather complex molecules in interstellar space.  As yet,
we do not know if the molecules were formed within an envelope of an AGB
star and survived ejection into interstellar space or if they were
incorporated onto dust grains and later evaporated from their surfaces
upon exposure to interstellar UV radiation.  It is likely that the 
true situation will be a mixture of both of these cases.

A number of molecules with transitions in the frequency range of the
SKA are of particular interest for astrobiology.  These include the
building block molecule for simple sugars such as ribose and
deoxyribose, Furan ( $C_{4}H_{4}O$ ; e.g. with a transition at 
10.6 GHz) \cite{Dickens2002}.  The same
authors detected c-C$_{2}$H$_{4}$O, one of the few cyclic molecules in
space.  They note that the presence of these molecular species in cold
dark clouds suggests that rather complex organic molecules may have
been present in the solar system before the planets formed, a first
step toward explaining the origin of life on the early Earth.

\section{Magnetic Fields}

Magnetic fields of AGB stars are now thought to be fairly strong from
observations of SiO masers \cite{Kemball1997}.  Depending on the exact
models used, the field strengths seem to be between 5-10 G at radii of 3 AU or
so.  Such strong magnetic fields will have obvious impacts on both the
molecular gas (Zeeman splitting for a number of species such as CCS
and SO) and possibly produce circularly polarized radio emission from
the star itself (analogous to the emission observed from the Sun).
Although requiring careful instrumental polarization characterization,
observations of these effects will provide confirmation of the
magnetic field strength implied by the SiO maser polarization
observations and further constraints on AGB stars themselves.
Potential movies of regions of magnetic field enhancements on the
surfaces of nearby AGB stars could be tracked with time, testing maser
observations of rotation.

\onecolumn
\section{Molecular Line Transitions Available to the SKA}
\setlongtables
\begin{tiny}
\begin{longtable}{lll|lll}
\caption{
\label{LOVASTABLE}
Molecular line transitions in the frequency range 700MHz to 30GHz detected at least once in the interstellar environment
\cite{LOVAS}. } \\
\hline
Frequency (MHz) & Formula & Quantum Numbers & Frequency (MHz) & Formula & Quantum Numbers \\
\hline
\endfirsthead
\hline
Frequency (MHz) & Formula & Quantum Numbers & Frequency (MHz) & Formula & Quantum Numbers \\
\hline
\endhead
\multicolumn{6}{r}{\it table continues...} \\
\hline
\endfoot
701.679 & CH & 23/2 J=3/2 F=2-2 & 19418.796 & c-C$_{3}$HD & 1(1,0)-1(0,1) F=1-0  \\
704.175  & CH & 23/2 J=3/2 F=2+-1- & 19426.679 & CH$_{2}$CHCN & 2(1,1)-1(1,0) F=2-1 \\
722.303  & CH & 23/2 J=3/2 F=1+-2- & 19427.851 & CH$_{2}$CHCN & 2(1,1)-1(1,0) F=3-2  \\
724.791 & CH & 23/2 J=3/2 F=1-1 & 19429.098 & CH$_{2}$CHCN & 2(1,1)-1(1,0) F=1-0  \\
834.285  & CH$_{3}$OH & 1(1,0)-1(1,1) A-+ & 19430.85 & unidentified &  \\
1065.076  & CH$_{3}$CHO & 1(1,0) - 1(1,1) A-+ & 19609.78 & unidentified &  \\
1371.722  & CH$_{2}$CHCN & 2(1,1)-2(1,2) F=1-1 & 19682.50 & unidentified &  \\
1371.797  & CH$_{2}$CHCN & 2(1,1)-2(1,2) F=3-3 & 19692.50 & unidentified &  \\
1371.934 & CH$_{2}$CHCN & 2(1,1)-2(1,2) F=2-2 & 19755.111 & HC$_{9}$N & 34-33  \\
1538.108 & NH$_{2}$CHO & 1(1,0)-1(1,1) F=1-1 & 19757.538 & NH$_{3}$ & 6(3)-6(3) \\
1538.676 & NH$_{2}$CHO & 1(1,0)-1(1,1) F=1-2 & 19771.50 & unidentified &  \\
1539.264 & NH$_{2}$CHO & 1(1,0)-1(1,1) F=2-1 & 19780.800 & CCCN & 2-1 J=5/2-3/2 F=5/2-3/2  \\
1539.527 & NH$_{2}$CHO & 1(1,0)-1(1,1) F=1-0 & 19780.826 & CCCN & 2-1 J=5/2-3/2 F=3/2-1/2  \\
1539.832 & NH$_{2}$CHO & 1(1,0)-1(1,1) F=2-2 & 19781.094 & CCCN & 2-1 J=5/2-3/2 F=7/2-5/2  \\
1540.998 & NH$_{2}$CHO & 1(1,0)-1(1,1) F=0-1 & 19799.951 & CCCN & 2-1 J=5/2-3/2 F=3/2-1/2  \\
1570.805 & NH$_{2}^{13}$CHO & 1(1,0)-1(1,1) F=2-2 & 19800.121 & CCCN & 2-1 J=5/2-3/2 F=5/2-3/2  \\
1584.274 & $^{18}$OH & 23/2 J=3/2 F=1-2 & 19838.346 & NH$_{3}$ & 5(1)-5(1) \\
1610.247 & CH$_{3}$OCHO & 1(1,0)-1(1,1) A & 19871.344 & HCCNC & 2-1  \\
1610.900 & CH$_{3}$OCHO & 1(1,0)-1(1,1) E & 19967.396 & CH$_{3}$OH & 2(1,1)-3(0,3) E \\
1612.2310 & OH & 23/2 J=3/2 F=1-2 & 19974.50 & unidentified &  \\
1624.518 & $^{17}$OH & 23/2 J=3/2 F,F1=7/2,4-7/2,4 & 20064.21 & unidentified &  \\
1626.161 & $^{17}$OH & 23/2 J=3/2 F,F1=9/2,4-9/2,4 & 20109.547 & CH$_{2}$CN & 1-0 3/2-1/2 5/2-3/2 5/2-5/2 \\
1637.564 & $^{18}$OH & 23/2 J=3/2 F=1-1 & 20115.77 & CH$_{2}$CN & 1-0 1/2-1/2 3/2-3/2 5/2-5/2  \\
1638.805 & HCOOH & 1(1,0)-1(1,1) & 20117.43 & CH$_{2}$CN & 1-0 3/2-1/2 5/2-3/2 3/2-1/2  \\
1639.503 & $^{18}$OH & 23/2 J-3/2 F=2-2 & 20118.014 & CH$_{2}$CN & 1-0 3/2-1/2 5/2-3/2 5/2-3/2  \\
1665.4018 & OH & 23/2 J=3/2 F=1-1 & 20118.16 & CH$_{2}$CN & 1-0 3/2-1/2 1/2-1/2 3/2-3/2  \\
1667.3590 & OH & 23/2 J=3/2 F=2-2 & 20119.606 & CH$_{2}$CN & 1-0 3/2-1/2 5/3-3/2 7/2-5/2  \\
1692.795 & $^{18}$OH & 23/2 J=3/2 F=2-1 & 20121.61 & CH$_{2}$CN & 1-0 3/2-1/2 3/2-3/2 3/2-3/2  \\
1720.5300 & OH & 23/2 J=3/2 F=2-1 & 20123.96 & CH$_{2}$CN & 1-0 3/2-1/2 1/2-1/2 3/2-3/2  \\
2661.61  & HC$_{5}$N & 1-0 F=1-1 & 20124.22 & CH$_{2}$CN & 1-0 1/2-1/2 3/2-1/2 3/2-1/2  \\
2662.87  & HC$_{5}$N & 1-0 F=2-1 & 20124.22 & CH$_{2}$CN & 1-0 3/2-1/2 3/2-3/2 1/2-1/2  \\
2664.76  & HC$_{5}$N & 1-0 F=0-1 & 20124.45 & CH$_{2}$CN & 1-0 3/2-1/2 3/2-1/2 3/2-3/2  \\
3139.404 & H$_{2}$CS & 2(1,1)-2(1,2) & 20124.49 & CH$_{2}$CN & 1-0 1/2-1/2 3/2-3/2 5/2-3/2  \\
3195.162 & CH$_{3}$CHO & 2(1,1) - 2(1,2) A-+ & 20126.031 & CH$_{2}$CN & 1-0 3/2-1/2 3/2-3/2 3/2-1/2  \\
3263.794 & CH & 21/2 J=1/2 F=0-1 & 20128.770 & CH$_{2}$CN & 1-0 1/2-1/2 3/2-1/2 3/2-3/2  \\
3335.481 & CH & 21/2 J=1/2 F=1-1 & 20139.76 & CH$_{2}$CN & 1-0 1/2-1/2 1/2-3/2 3/2-5/2  \\
3349.193 & CH & 21/2 J=1/2 F=1-0 & 20168.48 & unidentified & \\
4388.7786 & H$_{2}$C$^{18}$O & 1(1,0)-1(1,1) F=1-0 & 20171.089 & CH$_{3}$OH & 11(1,11)-10(2,8) A+  \\
4388.7960 & H$_{2}$C$^{18}$O & 1(1,0)-1(1,1) F=0-1 & 20203.31 & unidentified &  \\
4388.7963 & H$_{2}$C$^{18}$O & 1(1,0)-1(1,1) F=2-2 & 20209.209 & CH${_2}$CO & 1(0,1)-0(0,0) \\
4388.8011 & H$_{2}$C$^{18}$O & 1(1,0)-1(1,1) F=2-1 & 20281.00 & unidentified &  \\
4388.8035 & H$_{2}$C$^{18}$O & 1(1,0)-1(1,1) F=1-2 & 20303.946 & HC$_{7}$N & 18-17  \\
4388.8084 & H$_{2}$C$^{18}$O & 1(1,0)-1(1,1) F=1-1 & 20336.135 & HC$_{9}$N & 35-34  \\
4592.9563 & H$_{2}^{13}$CO & 1(1,0)-1(1,1)1/2,1/2-1/2,3/2 & 20357.226 & CH3C$_{4}$H & 5(1)-4(1) \\
4592.9738 & H$_{2}^{13}$CO & 1(1,0)-1(1,1)1/2,1/2-3/2,3/2 & 20357.423 & CH3C$_{4}$H & 5(0)-4(0)  \\
4592.9759 & H$_{2}^{13}$CO & 1(1,0)-1(1,1)3/2,1/2-1/2,3/2 & 20371.45 & NH$_{3}$ & 5(2)-5(2)  \\
4592.9857 & H$_{2}^{13}$CO & 1(1,0)-1(1,1)3/2,1/2-5/2,3/2 & 20460.01 & HDO & 3(2,1)-4(1,4)  \\
4592.9934 & H$_{2}^{13}$CO & 1(1,0)-1(1,1)3/2,1/2-3/2,3/2 & 20501.5 & unidentified &  \\
4593.0494 & H$_{2}^{13}$CO & 1(1,0)-1(1,1)1/2,1/2-1/2,1/2 & 20533.235 & unidentified &  \\
4593.0690 & H$_{2}^{13}$CO & 1(1,0)-1(1,1)3/2,1/2-1/2,1/2 & 20533.289 & C$_{8}$H & 23/2 17.5-16.5  \\
4593.0800 & H$_{2}^{13}$CO & 1(1,0)-1(1,1)1/2,1/2-3/2,1/2 & 20723.5 & unidentified &  \\
4593.0812 & H$_{2}^{13}$CO & 1(1,0)-1(1,1)1/2,3/2-1/2,3/2 & 20728.67 & unidentified &  \\
4593.0864 & H$_{2}^{13}$CO & 1(1,0)-1(1,1)3/2,3/2-1/2,3/2 & 20735.452 & NH$_{3}$ & 9(7)-9(7)  \\
4593.0865 & H$_{2}^{13}$CO & 1(1,0)-1(1,1)5/2,3/2-5/2,3/2 & 20765.80 & unidentified &  \\
4593.0942 & H$_{2}^{13}$CO & 1(1,0)-1(1,1)5/2,3/2-3/2,3/2 & 20790.00 & unidentified &  \\
4593.0961 & H$_{2}^{13}$CO & 1(1,0)-1(1,1)3/2,3/2-5/2,3/2 & 20792.563 & H$_{2}$CCC & 1(0,1)-0(0,0)  \\
4593.0985 & H$_{2}^{13}$CO & 1(1,0)-1(1,1)1/2,3/2-3/2,3/2 & 20792.872 & C$_{6}$H & 23/2 J=15/2-13/2 F=8-7 e  \\
4593.0994 & H$_{2}^{13}$CO & 1(1,0)-1(1,1)3/2,1/2-3/2,1/2 & 20792.945 & C$_{6}$H & 23/2 J=15/2-13/2 F=7-6 e  \\
4593.1039 & H$_{2}^{13}$CO & 1(1,0)-1(1,1)3/2,3/2-3/2,3/2 & 20794.444 & C$_{6}$H & 23/2 J=15/2-13/2 F=8-7 f  \\
4593.1741 & H$_{2}^{13}$CO & 1(1,0)-1(1,1)1/2,3/2-1/2,1/2 & 20794.512 & C$_{6}$H & 23/2 J=15/2-13/2 F=7-6 f  \\
4593.1795 & H$_{2}^{13}$CO & 1(1,0)-1(1,1)3/2,3/2-1/2,1/2 & 20804.830 & NH$_{3}$ & 7(5)-7(5)  \\
4593.2003 & H$_{2}^{13}$CO & 1(1,0)-1(1,1)5/2,3/2-3/2,1/2 & 20838.20 & unidentified &  \\
4593.2046 & H$_{2}^{13}$CO & 1(1,0)-1(1,1)1/2,3/2-3/2,1/2 & 20847.50 & unidentified &  \\
4593.2099 & H$_{2}^{13}$CO & 1(1,0)-1(1,1)3/2,3/2-3/2,1/2 & 20852.527 & NH$_{3}$ & 10(8)-10(8)  \\
4617.121 & NH$_{2}$CHO & 2(1,1)-2(1,2) F=2-2 & 20878.00 & unidentified &  \\
4618.967 & NH$_{2}$CHO & 2(1,1)-2(1,2) F=3-3 & 20908.848 & CH$_{3}$OH & 16(-4,13)-15(-5,10) E  \\
4619.993 & NH$_{2}$CHO & 2(1,1)-2(1,2) F=1-1 & 20917.157 & HC$_{9}$N & 36-35  \\
4660.242 & OH & 21/2 J=1/2 F=0-1 & 20970.658 & CH$_{3}$OH & 10(1,10)-11(,9) A+ t=1  \\
4750.656 & OH & 21/2 J=1/2 F=1-1 & 20994.617 & NH$_{3}$ & 6(4)-6(4)  \\
4765.562 & OH & 21/2 J=1/2 F=1-0 & 20999.79 & unidentified &  \\
4829.6412 & H${_2}$CO & 1(1,0)-1(1,1) F=1-0 & 21070.739 & NH$_{3}$ & 11(9)-11(9)  \\
4829.6587 & H${_2}$CO & 1(1,0)-1(1,1) F=0-1 & 21134.311 & NH$_{3}$ & 4(1)-4(1)  \\
4829.6594 & H${_2}$CO & 1(1,0)-1(1,1) F=2-2 & 21143.18 & unidentified &  \\
4829.6639 & H${_2}$CO & 1(1,0)-1(1,1) F=2-1 & 21231.00 & unidentified &  \\
4829.6664 & H${_2}$CO & 1(1,0)-1(1,1) F=1-2 & 21285.275 & NH$_{3}$ & 5(3)-5(3)  \\
4829.6710 & H${_2}$CO & 1(1,0)-1(1,1) F=1-1 & 21301.261 & HC$_{5}$N & 8-7  \\
4916.312 & HCOOH & 2(1,1)-2(1,2) & 21322.50 & unidentified &  \\
5005.3208 & CH$_{3}$OH & 3(1,2)-3(1,3) A-+ & 21431.932 & HC$_{7}$N & 19-18  \\
5289.015 & CH$_{2}$NH & 1(1,0)-1(1,1) F=0-1 & 21447.8 & unidentified &  \\
5289.678 & CH$_{2}$NH & 1(1,0)-1(1,1) F=1-0 & 21453.93 & unidentified &  \\
5289.813 & CH$_{2}$NH & 1(1,0)-1(1,1) F=2-2 & 21470.4 & unidentified &  \\
5290.614 & CH$_{2}$NH & 1(1,0)-1(1,1) F=2-1 & 21480.809 & C$_{5}$H & 21/2 J=9/2-7/2 F=5-4 e  \\
5290.879 & CH$_{2}$NH & 1(1,0)-1(1,1) F=1-2 & 21481.299 & C$_{5}$H & 21/2 J=9/2-7/2 F=4-3 e  \\
5291.680 & CH$_{2}$NH & 1(1,0)-1(1,1) F=1-1 & 21484.695 & C$_{5}$H & 21/2 J=9/2-7/2 F=5-4 f  \\
5324.058 & HC$_{5}$N & 2-1 F=2-2 & 21485.248 & C$_{5}$H & 21/2 J=9/2-7/2 F=4-3 f  \\
5324.270 & HC$_{5}$N & 2-1 F=1-0 & 21488.255 & H$_{2}$CCCCCC & 8(1,8)-7(1,7)  \\
5325.330 & HC$_{5}$N & 2-1 F=2-1 & 21498.182 & HC$_{9}$N & 37-36  \\
5325.421 & HC$_{5}$N & 2-1 F=3-2 & 21546.94 & unidentified &  \\
5327.451 & HC$_{5}$N & 2-1 F=1-1 & 21550.342 & CH$_{3}$OH & 12(2,11)-11(1,11) A+ t=1 \\
6016.746 & OH & 23/2 J=5/2 F=2-3 & 21569.5 & unidentified &  \\
6030.747 & OH & 23/2 J=5/2 F=2-2 & 21576.5 & unidentified &  \\
6035.092 & OH & 23/2 J-5/2 F=3-3 & 21582.6 & unidentified &  \\
6049.084 & OH & 23/2 J=5/2 F=3-2 & 21587.400 & c-C$_{3}$H$_{2}$ & 2(2,0)-2(1,1) \\
6278.628 & H$_{2}$CS & 3(1,2)-3(1,3) & 21592.1 & unidentified &  \\
6389.933 & CH$_{3}$CHO & 3(1,2) - 3(1,3) A-+ & 21595.8 & unidentified &  \\
6668.5192 & CH$_{3}$OH & 5(1,6)-6(0,6) A++ & 21598.4 & unidentified &  \\
7761.747 & OH & 21/2 J=3/2 F=1-1 & 21606.30 & unidentified &  \\
7820.125 & OH & 21/2 J=3/2 F=2-2 & 21615.5 & unidentified &  \\
7895.989 & HC$_{7}$N & 7-6 F=6-5 & 21703.3580 & NH$_{3}$ & 4(2)-4(2)  \\
7896.010 & HC$_{7}$N & 7-6 F=7-6 & 21715.8 & unidentified &  \\
7896.023 & HC$_{7}$N & 7-6 F=8-7 & 21930.476 & CC34S & 2,1-1,0  \\
7987.782 & HC$_{5}$N & 3-2 F=2-1 & 21980.5453 & HNCO & 1(0,1)-0(0,0) F=0-1  \\
7987.994 & HC$_{5}$N & 3-2 F=3-2 & 21981.4706 & HNCO & 1(0,1)-0(0,0) F=2-1  \\
7988.044 & HC$_{5}$N & 3-2 F=4-3 & 21982.0854 & HNCO & 1(0,1)-0(0,0) F=1-1  \\
8135.870 & OH & 21/2 J=5/2 F=2-2 & 22079.204 & HC$_{9}$N & 38-37  \\
8189.587 & OH & 21/2 J=5/2 F=3-3 & 22235.044 & H$_{2}$O & 6(1,6)-5(2,3) F=7-6  \\
8775.088 & CH$_{3}$NH$_{2}$ & 2(0,2)-1(0,1) F=1-0 Aa & 22235.077 & H$_{2}$O & 6(1,6)-5(2,3) F=6-5  \\
8777.442 & CH$_{3}$NH$_{2}$ & 2(0,2)-1(0,1) F=3-2 Aa & 22235.120 & H$_{2}$O & 6(1,6)-5(2,3) F=5-4  \\
8778.200 & CH$_{3}$NH$_{2}$ & 2(0,2)-1(0,1) F=2-2 Aa & 22235.253 & H$_{2}$O & 6(1,6)-5(2,3) F=6-6  \\
8778.260 & CH$_{3}$NH$_{2}$ & 2(0,2)-1(0,1) F=1-1 Aa & 22235.298 & H$_{2}$O & 6(1,6)-5(2,3) F=5-5  \\
8779.496 & CH$_{3}$NH$_{2}$ & 2(0,2)-1(0,1) F=2-1 Aa & 22258.173 & CCO & 2,1-1,0  \\
8815.814 & H$^{13}$CCCN & 1-0 F=1-1 & 22307.670 & HDO & 5(3,2)-5(3,3)  \\
8817.096 & H$^{13}$CCCN & 1-0 F=2-1 & 22344.030 & CCS & 2,1-1,0  \\
8819.019 & H$^{13}$CCCN & 1-0 F=0-1 & 22471.180 & HCOOH & 1(0,1)-0(0,0)  \\
9024.009 & HC$_{7}$N & 8-7 & 22559.915 & HC$_{7}$N & 20-19  \\
9058.447 & HC$^{13}$CCN & 1-0 F=1-1 & 22624.8892 & 15NH$_{3}$ & 1(1)-1(1) F,F1=1.5,1-1.3,1  \\
9059.318 & HCC$^{13}$CN & 1-0 F=1-1 & 22624.9331 & 15NH$_{3}$ & 1(1)-1(1) F,F1=1.5,1-0.8,1  \\
9059.736 & HC$^{13}$CCN & 1-0 F=2-1 & 22624.9410 & 15NH$_{3}$ & 1(1)-1(1) F,F1=0.5,1-0.8,1  \\
9060.6080 & HCC$^{13}$CN & 1-0 F=2-1 & 22624.9469 & 15NH$_{3}$ & 1(1)-1(1) F,F1=1.5,2-1.5,2  \\
9097.0346 & HCCCN & 1-0 F=1-1 & 22639.3 & unidentified &  \\
9098.3321 & HCCCN & 1-0 F=2-1 & 22644.3 & unidentified &  \\
9100.2727 & HCCCN & 1-0 F=0-1 & 22649.843 & 15NH$_{3}$ & 2(2)-2(2)  \\
9118.823 & CH$_{3}$OCH$_{3}$ & 2(0,2)-1(1,1) AA & 22653.022 & NH$_{3}$ & 5(4)-5(4) \\
9119.671 & CH$_{3}$OCH$_{3}$ & 2(0,2)-1(1,1) EE & 22660.225 & HC$_{9}$N & 39-38  \\
9120.509 & CH$_{3}$OCH$_{3}$ & 2(0,2)-1(1,1) AE & 22678.6 & unidentified &  \\
9120.527 & CH$_{3}$OCH$_{3}$ & 2(0,2)-1(1,1) EA & 22688.312 & NH$_{3}$ & 4(3)-4(3)  \\
9235.119 & NH$_{2}$CHO & 3(1,2)-3(1,3) F=3-3 & 22732.429 & NH$_{3}$ & 6(5)-6(5)  \\
9237.034 & NH$_{2}$CHO & 3(1,2)-3(1,3) F=4-4 & 22789.421 & 15NH$_{3}$ & 3(3)-3(3)  \\
9237.704 & NH$_{2}$CHO & 3(1,2)-3(1,3) F=2-2 & 22827.741 & CH$_{3}$OCHO & 2(1,2)-1(1,1) E  \\
9486.71 & unidentified & & 22828.134 & CH$_{3}$OCHO & 2(1,2)-1(1,1) A  \\
9493.061 & C$_{4}$H & 3/2-1/2 F=1-0 & 22834.1851 & NH$_{3}$ & 3(2)-3(2)  \\
9496.4 & unidentified & & 22878.949 & DC5N & 9-8  \\
9497.616 & C$_{4}$H & 3/2-1/2 F=2-1 & 22924.940 & NH$_{3}$ & 7(6)-7(6)  \\
9508.005 & C$_{4}$H & 3/2-1/2 F=1-1 & 23046.0158 & 15NH$_{3}$ & 4(4)-4(4)  \\
9547.953 & C$_{4}$H & 1/2-1/2 F=1-0 & 23098.8190 & NH$_{3}$ & 2(1)-2(1)  \\
9551.717 & C$_{4}$H & 1/2-1/2 F=0-1 & 23121.024 & CH$_{3}$OH & 9(2,7)-10(1,10) A+ \\
9562.904 & C$_{4}$H & 1/2-1/2 F=1-1 & 23122.983 & CCCS & 4-3  \\
9703.508 & C$_{6}$H & 23/2 J=3.5-2.5 F=4-3 e & 23142.2 & unidentified & \\
9703.600 & C$_{6}$H & 23/2 J=3.5-2.5 F=3-2 e & 23228.0 & unidentified &  \\
9703.835 & C$_{6}$H & 23/2 J=3.5-2.5 F=4-3 f & 23232.238 & NH$_{3}$ & 8(7)-8(7)  \\
9703.936 & C$_{6}$H & 23/2 J=3.5-2.5 F=3-2 f & 23241.246 & HC$_{9}$N & 40-39  \\
9877.606 & HC$_{9}$N & 17-16 & 23421.9823 & 15NH$_{3}$ & 5(5)-5(5)  \\
9885.89  & CCCN & 1-0 J=3/2-1/2 F=5/2-3/2 & 23444.778 & CH$_{3}$OH & 10(1,9)-9(2,8) A-  \\
9936.202 & CH$_{3}$OH & 9(-1,9)-8(-2,7) E & 23565.160 & C$_{6}$H & 23/2 J=17/2-15/2 F=9-8 e  \\
9978.686 & CH$_{3}$OH & 4(3,2)-5(2,3) E & 23565.226 & C$_{6}$H & 23/2 J=17/2-15/2 F=8-7 e  \\
10058.257 & CH$_{3}$OH & 4(3,1)-5(2,4) E & 23567.169 & C$_{6}$H & 23/2 J=17/2-15/2 F=9-8 f  \\
10152.008 & HC$_{7}$N & 9-8 & 23567.238 & C$_{6}$H & 23/2 J=17/2-15/2 F=8-7 f  \\
10278.246 & HDO & 2(2,0)-2(2,1) & 23600.242 & SiC2 & 1(0,1)-0(0,0)  \\
10458.639 & HC$_{9}$N & 18-17 & 23657.471 & NH$_{3}$ & 9(8)-9(8)  \\
10463.962 & H$_{2}$CS & 4(1,3)-4(1,4) & 23687.898 & HC$_{7}$N & 21-20  \\
10648.419 & CH$_{3}$CHO & 4(1,3) - 4(1,4) A-+ & 23692.9265 & NH$_{3}$ & 1(1)-1(1) F,F1=1/2,1-1/2,0  \\
10650.563 & HC$_{5}$N & 4-3 F=3-2 & 23692.9688 & NH$_{3}$ & 1(1)-1(1) F,F1=3/2,1-1/2,0  \\
10650.654 & HC$_{5}$N & 4-3 F=4-3 & 23693.8722 & NH$_{3}$ & 1(1)-1(1) F,F1=1/2,1-3/2,2  \\
10650.686 & HC$_{5}$N & 4-3 F=5-4 & 23693.9051 & NH$_{3}$ & 1(1)-1(1) F,F1=3/2,1-5/2,2  \\
11119.445 & CCS & 1,0-0,1 & 23693.9145 & NH$_{3}$ & 1(1)-1(1) F,F1=3/2,1-3/2,2  \\
11280.006 & HC$_{7}$N & 10-9 & 23694.4591 & NH$_{3}$ & 1(1)-1(1) F,F1=1/2,1-1/2,1  \\
11561.513 & CCCS & 2-1 & 23694.4700 & NH$_{3}$ & 1(1)-1(1) F,F1=1/2,1-3/2,1  \\
12162.979 & OCS & 1-0 & 23694.4709 & NH$_{3}$ & 1(1)-1(1) F,F1=3/2,2-5/2,2  \\
12178.593 & CH$_{3}$OH & 2(0,2)-3(-1,3) E & 23694.4803 & NH$_{3}$ & 1(1)-1(1) F,F1=3/2,2-3/2,2 \\
12408.003 & HC$_{7}$N & 11-10 & 23694.5014 & NH$_{3}$ & 1(1)-1(1) F,F1=3/2,1-1/2,1  \\
12782.769 & HC$_{9}$N & 22-21 & 23694.5060 & NH$_{3}$ & 1(1)-1(1) F,F1=5/2,2-5/2,2  \\
12848.48 & unidentified & & 23694.5123 & NH$_{3}$ & 1(1)-1(1) F,F1=3/2,1-3/2,1  \\
12848.731 & HC$^{11}$N & 38-37 & 23694.5153 & NH$_{3}$ & 1(1)-1(1) F,F1=5/2,2-3/2,2  \\
13043.814 & SO & 1(2)-1(1) & 23695.0672 & NH$_{3}$ & 1(1)-1(1) F,F1=3/2,2-3/2,1  \\
13116.451 & unidentified & & 23695.0782 & NH$_{3}$ & 1(1)-1(1) F,F1=3/2,2-3/2,1  \\
13116.569 & unidentified & & 23695.1132 & NH$_{3}$ & 1(1)-1(1) F,F1=5/2,2-3/2,1  \\
13186.46 & unidentified & & 23696.0297 & NH$_{3}$ & 1(1)-1(1) F,F1=1/2,0-1/2,1  \\
13186.853 & HC$^{11}$N & 39-38 & 23696.0406 & NH$_{3}$ & 1(1)-1(1) F,F1=1/2,0-3/2,1  \\
13186.98 & unidentified & & 23697.9 & unidentified &  \\
13313.312 & HC$_{5}$N & 5-4 & 23718.325 & HC13CCCCN & 9-8  \\
13363.801 & HC$_{9}$N & 23-22 & 23720.575 & NH$_{3}$ & 2(2)-2(2) F1=1-2  \\
13434.596 & OH & 23/2 J=7/2 F=3-3 & 23721.336 & NH$_{3}$ & 2(2)-2(2) F1=3-2  \\
13441.4173 & OH & 23/2 J=7/2 F=4-4 & 23722.6323 & NH$_{3}$ & 2(2)-2(2) F1=2-2  \\
13535.998 & HC$_{7}$N & 12-11 & 23722.6336 & NH$_{3}$ & 2(2)-2(2) F1=3-3  \\
13778.804 & H$_{2}^{13}$CO & 2(1,1)-2(1,2) & 23722.6344 & NH$_{3}$ & 2(2)-2(2) F1=1-1  \\
13880.54 & unidentified & & 23723.929 & NH$_{3}$ & 2(2)-2(2) F1=2-3  \\
13944.832 & HC$_{9}$N & 24-23 & 23724.691 & NH$_{3}$ & 2(2)-2(2) F1=2-1  \\
14488.4589 & H${_2}$CO & 2(1,1)-2(1,2) F=1-1 & 23727.162 & HCCCC13CN & 9-8  \\
14488.4712 & H${_2}$CO & 2(1,1)-2(1,2) F=1-2 & 23804.5 & unidentified &  \\
14488.4801 & H${_2}$CO & 2(1,1)-2(1,2) F=3-3 & 23811.0 & unidentified &  \\
14488.4899 & H${_2}$CO & 2(1,1)-2(1,2) F=2-2 & 23817.6153 & OH & 23/2 J=9/2 F=4-4  \\
14525.862 & HC$_{9}$N & 25-24 & 23822.265 & HC$_{9}$N & 41-40  \\
14663.993 & HC$_{7}$N & 13-12 & 23826.6211 & OH & 23/2 J=9/2 F=5-5  \\
14782.212 & 13CH$_{3}$OH & 2(0,2)-3(-1,3) E & 23867.805 & NH$_{3}$ & 3(3)-3(3) F1=2-3  \\
14812.002 & c-C$_{3}$H & 1(1,0)-1(1,1) J=3/2-1/2 F=2-1 & 23868.450 & NH$_{3}$ & 3(3)-3(3) F1=4-3  \\
14877.671 & c-C$_{3}$H & 1(1,0)-1(1,1) J=3/2-3/2 F=2-1 & 23870.1279 & NH$_{3}$ & 3(3)-3(3) F1=3-3  \\
14893.050 & c-C$_{3}$H & 1(1,0)-1(1,1) J=3/2-3/2 F=2-2 & 23870.1296 & NH$_{3}$ & 3(3)-3(3) F1=4-4  \\
14895.243 & c-C$_{3}$H & 1(1,0)-1(1,1) J=3/2-3/2 F=1-1 & 23870.1302 & NH$_{3}$ & 3(3)-3(3) F1=2-2  \\
15106.892 & HC$_{9}$N & 26-25 & 23871.807 & NH$_{3}$ & 3(3)-3(3) F1=3-4  \\
15248.225 & C$_{6}$H & 23/2 J=11/2-9/2 F=6-5 f & 23872.453 & NH$_{3}$ & 3(3)-3(3) F1=3-2  \\
15248.359 & C$_{6}$H & 23/2 J=11/2-9/2 F=5-4 f & 23922.3132 & 15NH$_{3}$ & 6(6)-6(6)  \\
15249.064 & C$_{6}$H & 23/2 J=11/2-9/2 F=6-5 e & 23939.089 & HCC13CCCN & 9-8  \\
15249.198 & C$_{6}$H & 23/2 J=11/2-9/2 F=5-4 e & 23941.99  & HCCC13CCN & 9-8  \\
15687.921 & HC$_{9}$N & 27-26 & 23959.5 & unidentified &  \\
15791.986 & HC$_{7}$N & 14-13 & 23963.901 & HC$_{5}$N & 9-8  \\
15975.966 & HC$_{5}$N & 6-5 & 23987.5 & unidentified &  \\
16268.950 & HC$_{9}$N & 28-27 & 23990.2 & unidentified &  \\
16849.979 & HC$_{9}$N & 29-28 & 23996.7 & unidentified &  \\
16886.312 & DCCCN & 2-1 F=2-1 & 24004.5 & unidentified &  \\
16886.405 & DC$_{3}$N & 2-1 F=3-2 & 24023.2 & unidentified &  \\
16919.979 & HC$_{7}$N & 15-14 & 24037.1 & unidentified &  \\
17091.742 & CH$_{3}$CCH & 1(0)-0(0) & 24048.5 & unidentified &  \\
17342.256 & CCCS & 3-2 & 24139.4169 & NH$_{3}$ & 4(4)-4(4)  \\
17431.006 & HC$_{9}$N & 30-59 & 24205.287 & NH$_{3}$ & 10(9)-10(9) \\
17632.685 & H$^{13}$CCCN & 2-1 F=2-2 & 24296.491 & CH$_{3}$OCHO & 2(0,2)-1(0,1) E  \\
17633.844 & H$^{13}$CCCN & 2-1 F=3-2 &  24298.481 & CH$_{3}$OCHO & 2(0,2)-1(0,1) A  \\
17647.479 & C$_{4}$D & 5/2-3/2 F=5/2-3/2 & 24325.927 & OCS & 2-1  \\
17647.526 & C$_{4}$D & 5/2-3/2 F=3/2-1/2 & 24375.2 & unidentified & \\
17647.716 & C$_{4}$D & 5/2-3/2 F=7/2-5/2 & 24428.652 & CH3C$_{4}$H & 6(1)-5(1)  \\
17666.995 & HCCC15N & 2-1 & 24428.886 & CH3C$_{4}$H & 6(0)-5(0)  \\
17683.961 & C$_{4}$D & 3/2-1/2 F=5/2-3/2 & 24532.9887 & NH$_{3}$ & 5(5)-5(5)  \\
17684.662 & C$_{4}$D & 3/2-1/2 F=3/2-1/2 & 24788.541 & CH3CCCN & 6(1)-5(1)  \\
17736.75 & unidentified & & 24788.780 & CH3CCCN & 6(0)-5(0)  \\
17788.570 & H$_{2}$CCCC & 2(1,2)-1(1,1) & 24815.878 & HC$_{7}$N & 22-21  \\
17863.803 & H$_{2}$CCCC & 2(0,2)-1(0,1) & 24928.715 & CH$_{3}$OH & 3(2,1)-3(1,2) E  \\
17937.956 & H$_{2}$CCCC & 2(1,1)-1(1,0) & 24933.468 & CH$_{3}$OH & 4(2,2)-4(1,3) E  \\
17945.85 & unidentified & & 24934.382 & CH$_{3}$OH & 2(2,0)-2(1,1) E  \\
17951.95 & unidentified & & 24959.079 & CH$_{3}$OH & 5(2,3)-5(1,4) E  \\
17965.09 & unidentified & & 24984.302 & HC$_{9}$N & 43-42  \\
17974.01 & unidentified & & 24991.19  & SiC2 & 8(2,6)-8(2,7)  \\
18012.033 & HC$_{9}$N & 31-30 & 25018.123 & CH$_{3}$OH & 6(2,4)-6(1,5) E  \\
18012.46 & unidentified & & 25023.792 & NH$_{2}$D & 4(1,4)-4(0,4)  \\
18017.337 & NH$_{3}$ & 7(3)-7(3) & 25056.025 & NH$_{3}$ & 6(6)-6(6)  \\
18020.574 & C$_{6}$H & 23/2 J=6.5-5.5 F=7-6 e & 25124.872 & CH$_{3}$OH & 7(2,5)-7(1,6) E  \\
18020.644 & C$_{6}$H & 23/2 J=6.5-5.5 F=6-5 e & 25249.938 & C5N & 21/2 N=9-8 J=9.5-8.5 \\
18021.752 & C$_{6}$H & 23/2 J=6.5-5.5 F=7-6 f & 25260.649 & C5N & 21/2 N=9-8 J=8.5-7.5  \\
18021.818 & C$_{6}$H & 23/2 J=6.5-5.5 F=6-5 f & 25294.417 & CH$_{3}$OH & 8(2,6)-8(1,7) E  \\
18021.86 & unidentified & & 25329.441 & DC$_{3}$N & 3-2  \\
18047.969 & HC$_{7}$N & 16-15 & 25421.036 & DC5N & 10-9  \\
18119.029 & HC$^{13}$CCN & 2-1 F=2-1 & 25541.398 & CH$_{3}$OH & 9(2,7)-9(1,8) E  \\
18120.773 & HCC$^{13}$CN & 2-1 F=2-1 & 25715.182 & NH$_{3}$ & 7(7)-7(7)  \\
18120.865 & HCC$^{13}$CN & 2-1 F=3-2 & 25878.266 & CH$_{3}$OH & 10(2,8)-10(1,9) E  \\
18154.884 & SiS & 1-0 & 25911.017 & CCS & 2,2-1,1  \\
18186.652 & C$_{8}$H & 23/2 15.5-15.5 e & 25943.855 & HC$_{7}$N & 23-22 \\
18186.782 & C$_{8}$H & 23/2 15.5-15.5 f & 26337.414 & C$_{6}$H & 23/2 J=19/2-17/2 F=10-9 f  \\
18194.9206 & HCCCN & 2-1 F=2-2 & 26337.463 & C$_{6}$H & 23/2 J=19/2-17/2 F=9-8 f  \\
18195.3176 & HCCCN & 2-1 F=1-0 & 26339.924 & C$_{6}$H & 23/2 J=19/2-17/2 F=10-9 e  \\
18196.2183 & HCCCN & 2-1 F=2-1 & 26339.973 & C$_{6}$H & 23/2 J=19/2-17/2 F=9-8 e  \\
18196.3119 & HCCCN & 2-1 F=3-2 & 26363.491 & HCCCC$^{13}$CN & 10-9  \\
18197.078 & HCCCN & 2-1 F=1-2 & 26450.598 & H$^{13}$CCCN & 3-2  \\
18198.3756 & HCCCN & 2-1 F=1-1 & 26500.462 & HCCC$^{15}$N & 3-2  \\
18222.65 & unidentified & & 26518.981 & NH$_{3}$ & 8(8)-8(8)  \\
18285.434 & NH$_{3}$ & 10(7)-10(7) & 26602.181 & HCCC$^{13}$CCN & 10-9  \\
18294.20 & unidentified & & 26626.533 & HC$_{5}$N & 10-9  \\
18299.5 & unidentified & & 26682.814 & H$_{2}$CCCC & 3(1,3)-2(1,2)  \\
18306.3 & unidentified & & 26795.635 & H$_{2}$CCCC & 3(0,3)-2(0,2)  \\
18320.7 & unidentified & & 26847.205 & CH$_{3}$OH & 12(2,10)-12(1,11) E \\
18343.144 & c-C$_{3}$H$_{2}$ & 1(1,0)-1(0,1) & 26906.891 & H$_{2}$CCCC & 3(1,2)-2(1,1)  \\
18360.50 & unidentified & & 27071.824 & HC$_{7}$N & 24-23  \\
18363.045 & unidentified & & 27084.348 & c-C$_{3}$H$_{2}$ & 3(3,0)-3(2,1)  \\
18363.142 & unidentified & & 27178.511 & HC$^{13}$CCN & 3-2  \\
18363.306 & unidentified & & 27181.127 & HCC$^{13}$CN & 3-2  \\
18363.406 & unidentified & & 27292.903 & HCCCN & 3-2 F=3-3 \\
18368.0 & unidentified & & 27294.078 & HCCCN & 3-2 F=2-1  \\
18379.6 & unidentified & & 27294.295 & HCCCN & 3-2 F=3-2  \\
18383.3 & unidentified & & 27294.347 & HCCCN & 3-2 F=4-3  \\
18391.562 & NH$_{3}$ & 6(1)-6(1) & 27296.235 & HCCCN & 3-2 F=2-2  \\
18396.7252 & CH$_{3}$CN & 1(0)-0(0) F=1-1 & 27472.501 & CH$_{3}$OH & 13(2,11)-13(1,12) E  \\
18397.9965 & CH$_{3}$CN & 1(0)-0(0) F=2-1 & 27477.943 & NH$_{3}$ & 9(9)-9(9)  \\
18399.8924 & CH$_{3}$CN & 1(0)-0(0) F=0-1 & 28009.975 & HNCCC & 3-2  \\
18413.822 & c-H13CCCH & 1(1,0)-1(0,1) & 28169.437 & CH$_{3}$OH & 14(2,12)-14(1,13) E  \\
18422.00 & unidentified & & 28199.804 & HC$_{7}$N & 25-24  \\
18485.07 & unidentified & & 28199.805 & HC$_{7}$N & 25-24  \\
18494.1 & CH3SH & 18(2)-17(3) A+ & 28316.031 & CH$_{3}$OH & 4(0,4)-3(1,2) E  \\
18499.390 & NH$_{3}$ & 9(6)-9(6) & 28440.980 & CH$_{2}$CHCN & 3(0,3)-2(0,2)  \\
18513.316 & CH$_{2}$CHCN & 2(1,2)-1(1,1) F=3-2 & 28470.391 & HC$_{9}$N & 49-48  \\
18586.06 & unidentified & & 28532.31 & C$_{4}$H & 7/2-5/2 F=3-2  \\
18593.060 & HC$_{9}$N & 32-31 & 28532.46 & C$_{4}$H & 7/2-5/2 F=4-3  \\
18638.616 & HC$_{5}$N & 7-6 & 28542.284 & C$_{4}$H & J=5/2-5/2 F=3-3  \\
18650.308 & HCCCHO & 2(0,2)-1(0,1) & 28571.37 & C$_{4}$H & 5/2-3/2 F=3-2  \\
18673.312 & HNCCC & 2-1 & 28571.53 & C$_{4}$H & 5/2-3/2 F=2-1  \\
18698.16 & unidentified & & 28604.737 & NH$_{3}$ & 10(10)-10(10)  \\
18729.12 & unidentified & & 28903.688 & CCCS & 5-4  \\
18793.92 & unidentified & & 28905.787 & CH$_{3}$OH & 15(2,13)-12(1,14) E  \\
18802.235 & H$_{2}$CCCCCC & 7(1,7)-6(1,6) & 28919.931 & CH$_{3}$CCCN & 7(1)-6(1)  \\
18807.888 & NH$_{2}$D & 3(1,3)-3(0,3) & 28920.209 & CH$_{3}$CCCN & 7(0)-6(0)  \\
18808.507 & NH$_{3}$ & 8(5)-8(5) & 28969.954 & CH$_{3}$OH & 8(2,7)-9(1,8) A--  \\
18817.66 & unidentified & & 28974.781 & H${_2}$CO & 3(1,2)-3(1,3) F=2-2  \\
18864.65 & unidentified & & 28974.804 & H${_2}$CO & 3(1,2)-3(1,3) F=4-4  \\
18884.695 & NH$_{3}$ & 6(2)-6(2) & 28974.814 & H${_2}$CO & 3(1,2)-3(1,3) F=3-3  \\
18907.54 & unidentified & & 28999.814 & HCCCC$^{13}$CN & 11-10  \\
18918.50 & unidentified & & 29051.403 & HC$_{9}$N & 50-49  \\
18961.79 & unidentified & & 29109.644 & C$_{6}$H & 23/2 J=21/2-19/2 F=11-10 f  \\
18965.588 & CH$_{2}$CHCN & 2(0,2)-1(0,1) F=1-0 & 29109.66  & C$_{6}$H & 23/2 J=21/2-19/2 f  \\
18966.535 & CH$_{2}$CHCN & 2(0,2)-1(0,1) F=2-1 & 29109.686 & C$_{6}$H & 23/2 J=21/2-19/2 F=10-9 f  \\
18966.616 & CH$_{2}$CHCN & 2(0,2)-1(0,1) F=3-2 & 29112.709 & C$_{6}$H & 23/2 J=21/2-19/2 F=11-10 f  \\
18968.48 & unidentified & & 29112.73  & C$_{6}$H & 23/2 J=21/2-19/2 e  \\
18986.20 & unidentified & & 29112.750 & C$_{6}$H & 23/2 J=21/2-19/2 F=10-9 f  \\
19014.7204 & C$_{4}$H & 5/2-3/2 F=2-1 & 29138.877 & CH$_{2}$CHCN & 3(1,2)-2(1,1) F=3-2  \\
19015.1435 & C$_{4}$H & 5/2-3/2 F=3-2 & 29139.215 & CH$_{2}$CHCN & 3(1,2)-2(1,1) F=4-3, 2-1  \\
19025.107 & C$_{4}$H & 5/2-3/2 F=2-2 & 29258.834 & HCC$^{13}$CCCN & 11-10  \\
19039.50 & unidentified & & 29289.159 & HC$_{5}$N & 11-10  \\
19043.0 & unidentified & & 29304.09  & C$_{6}$H & 21/2 J=21/2-19/2 e  \\
19044.760 & C$_{4}$H & 3/2-1/2 F=1-1 & 29310.5 & unidentified &  \\
19054.4762 & C$_{4}$H & 3/2-1/2 F=2-1 & 29327.776 & HC$_{7}$N & 26-25  \\
19055.9468 & C$_{4}$H & 3/2-1/2 F=1-0 & 29332.45  & C$_{6}$H & 21/2 J=21/2-19/2 f  \\
19099.656 & C$_{4}$H & 3/2-3/2 F=1-1 & 29333.3 & unidentified &  \\
19119.764 & C$_{4}$H & J=3/2-3/2 F=2-2 & 29337.57  & HC$_{5}$N & 11-10 v11=1 =1c  \\
19174.086 & HC$_{9}$N & 33-32 & 29342.0 & unidentified &  \\
19175.958 & HC$_{7}$N & 17-16 & 29353.8 & unidentified &  \\
19218.465 & NH$_{3}$ & 7(4)-7(4) & 29363.15  & HC$_{5}$N & 11-10 v11=1 =1d \\
19243.521 & CCCO & 2-1 & 29365.0 & unidentified &  \\
19262.140 & CH$_{3}$CHO & 1(0,1) - 0(0,0) E & 29477.704 & CCS & 2,3-1,2  \\
19265.137 & CH$_{3}$CHO & 1(0,1) - 0(0,0) A++ & 29632.406 & HC$_{9}$N & 51-50  \\
19316.70 & unidentified & & 29632.413 & HC$_{9}$N & 51-50  \\
19325.20 & unidentified & & 29636.920 & CH$_{3}$OH & 16(2,14)-12(1,15) E \\
19336.10 & unidentified & & 29676.14 & CCCN & 3-2 J=7/2-5/2 F=7/2-5/2  \\
19361.50 & unidentified & & 29676.28 & CCCN & 3-2 J=7/2-5/2 F=9/2-7/2  \\
19418.661 & c-C$_{3}$HD & 1(1,0)-1(0,1) F=1-1 & 29678.882 & $^{34}$~SO & 1(0)-0(1)  \\
19418.686 & c-C$_{3}$HD & 1(1,0)-1(0,1) F=2-1 & 29694.99 & CCCN & 3-2 J=5/2-3/2 F=3/2-1/2  \\
19418.712 & c-C$_{3}$HD & 1(1,0)-1(0,1) F=1-2 & 29695.14 & CCCN & 3-2 J=5/2-3/2 F=7/2-5/2  \\
19418.724 & c-C$_{3}$HD & 1(1,0)-1(0,1) F=0-1 & 29806.963 & HCCNC & 3-2  \\
19418.740 & c-C$_{3}$HD & 1(1,0)-1(0,1) F=2-2 & 29914.486 & NH$_{3}$ & 11(11)-11(11)  \\
\hline
\end{longtable}
\end{tiny}
\end{document}